\documentclass[twocolumn,showpacs,superscriptaddress,pra]{revtex4}
\usepackage{amssymb}
\usepackage{amsfonts}
\usepackage{amsmath}
\usepackage{graphicx}
\usepackage{dcolumn}
\usepackage{bm}

\begin{document}

\title{On quantum control limited by quantum decoherence }
\author{Fei \surname{Xue}}
\email{Echoxue@itp.edu.cn}
\affiliation{Institute of Theoretical Physics, Chinese Academy of Sciences, 100080, China}
\author{S.X. \surname{Yu}}
\affiliation{Department of Modern Physics, University of Science and Technology of China,
Hefei, Anhui 230026, China}
\author{C.P. \surname{Sun}}
\email{suncp@itp.edu.cn}
\affiliation{Institute of Theoretical Physics, Chinese Academy of Sciences, 100080, China}
\date{July 22, 2005}

\begin{abstract}
We describe the quantum controllability under the influences of the quantum
decoherence induced by the quantum control itself. It is shown that, when
the controller is considered as a quantum system, it will entangle with its
controlled system and then cause the quantum decoherence in the controlled
system. In competition with this induced decoherence, the controllability
will be limited by some uncertainty relation in a well-armed quantum control
process. In association with the phase uncertainty and the standard quantum
limit, a general model is studied to demonstrate the possibility of
realizing a decoherence-free quantum control with a finite energy within a
finite time. It is also shown that if the operations of quantum control are
to be determined by the initial state of the controller, then due to the
decoherence which results from the quantum control itself, there exists a
low bound for the quantum controllability.
\end{abstract}

 \maketitle

\section{Introduction}

Generally people can utilize an external field to manipulate the time
evolution of a quantum system from an arbitrary initial state to reach any
wanted target state. If the external field is classical and can be
artificially controlled to be time-dependent, then we refer this kind of
manipulation as a classical control \cite{q-control}. In quantum
computations \cite{q-inform}, the quantum logic gate operations can be
regarded as classical controls in most cases where the controller is
essentially classical and the control can be turned on or off classically at
certain instants.

In this paper we consider the quantum control, in which the controller is
quantized and obeys the laws of quantum mechanics. It is shown that the back
action of the controlled system should be considered, which may have a
negative side-effect on the controllability. There are two motivations for
our investigations.

Firstly, it is exciting to explore the finiteness of human being's abilities
to control the nature, and a ``down-to-earth" starting point for this
exploration in physics should be a concrete model even though it is
oversimplified. With some reasonable models one could demonstrate how the
fundamental laws of physics impose the limits on the controllability in
principle. These refer to some basic issues in physics, such as the energy
bound, the basic precision of measurement (or standard quantum limit (SQL)%
\cite{SQL}). It is emphasized that the quantum decoherence may result from
\emph{the control itself} when the controller is essentially considered as a
quantum subsystem.

Secondly, though the physical implementation of quantum computation seems to
be difficult, the huge power of quantum computation has been demonstrated by
some quantum algorithms in principle. The limit of quantum control can bring
a physical limit to quantum computation architecture since it is based on
complete quantum blocks including the controller. Lloyd discussed how the
physical constants impose a limit on the power and the memory in the quantum
computer \cite{Lloyd2000}, while Ozawa and Gea-Banacloche \cite%
{Ozawa2002,Julio2002b} considered the conservation law and the minimum
energy requirement for quantum computation respectively. Our present study
can also be regarded as a part of the growing body of the explorations in
this direction.

In Sec. \ref{sec:single}, we start with a model with a single mode field as
a controller and a two-level system (qubit) as the controlled system. We
found that it is possible to implement some phase gate controls without
inducing decoherence to the controlled system. However, the single mode
example is far from practical cases, and thus we further study the quantum
control in a more general case in Sec. \ref{sec:general model}. In Sec. \ref%
{sec:uphase} the control induced decoherence is explained as a phase
uncertainty by associating it with the SQL. In Sec. \ref{sec:inequality} the
obtained results is highlighted as the complementarity of the
controllability and the control induced decoherence. An inequality similar
to the Heisenberg uncertainty relation is presented as the accurate bound of
quantum gates under the quantum control.

\section{An exactly soluble model for the quantum control}

\label{sec:single}

To have a clear picture about the quantum control, let us first start with a
simple model. The total system that we concern is closed, which consists of
the controller $C$ with the Hamiltonian $H_{c}$ and the controlled system $Q$
with the Hamiltonian $H_{q}$. The system is in the initial states $|\psi
_{c}(0,R)\rangle $ and $|\psi _{q}(0)\rangle =\sum_{n}c_{n}|n\rangle $
respectively, where $R$ represents the controlling parameters. For a given
target state $|\psi _{t}\rangle $ of $Q$, the quantum control is described
as a factorized evolution
\begin{equation}
|\psi _{q}(0)\rangle \otimes |\psi _{c}(0,R)\rangle \rightarrow |\psi
_{q}(T)\rangle \otimes |\psi _{c}(T)\rangle
\end{equation}
of the total system driven by the interaction Hamiltonian $H_{qc}$ within
the time duration $(0,T)$. If one could choose an appropriate initial state
and the corresponding parameters $R$ such that the partial wave function $%
|\psi _{q}(T)\rangle \equiv U_{q}(T)|\psi _{q}(0)\rangle $ is just the
target one $|\psi _{t}\rangle $, where a global phase difference is allowed,
then we could say that an ideal quantum control is realized. Usually $%
U_{q}(T)$ defines a quantum logic operation in quantum computation.

We now consider an exactly soluble example, where the controlled system is a
qubit with two basis states $|0\rangle $ and $|1\rangle $ and the controller
is a single mode boson field with free Hamiltonian $H_{c}=\hbar \omega
a^{\dag }a$, here $a^{\dag }$ ($a$) is the creation (annihilation) operator.
The interaction
\begin{equation}
H_{qc}=|1\rangle \langle 1|\otimes V\equiv 1\rangle \langle 1|\otimes
(ga+g^{\ast }a^{\dag })
\end{equation}
between them is of non-demolition\cite{SQL}, i.e., $[H_{qc},H_{c}]\neq 0$
and $[H_{qc},H_{q}]=0$. Since $H_{q}$ is conserved during the evolution we
take $H_{q}=0$ without loss of the generality. In the interaction picture
the time-dependent potential
\begin{equation}
V_{I}(t)=gae^{-i\omega t}+h.c
\end{equation}
acts only on the state $|1\rangle $, but not on $|0\rangle $. This
Hamiltonian originates from the atom-field system in the large detuning
limit, but the problem is greatly simplified for convenience \cite{q-opt}.

Now we explore the possibility of automatically creating a phase gate
operation
\begin{equation}
|\psi _{q}(0)\rangle =c_{0}|0\rangle +c_{1}|1\rangle \rightarrow |\psi
_{q}(t)\rangle =c_{0}|0\rangle +c_{1}e^{i\phi }|1\rangle  \label{eqn:4}
\end{equation}
driven by $H_{qc}$. Essentially, the phase gate operation is supposed to
generate a relative phase $\phi $ between $|0\rangle $ and $|1\rangle $ and
the total system experiences a factorized evolution
\begin{equation}
(c_{0}|0\rangle +c_{1}|1\rangle )\otimes |\psi _{c}(0)\rangle \rightarrow
(c_{0}|0\rangle +c_{1}e^{i\phi }|1\rangle )\otimes |\psi _{c}(T)\rangle .
\label{eqn:eqn3}
\end{equation}
We will show that, only a class of phase gates with special phases depending
on the global parameters, such as the coupling coefficients $g$ and the gate
operation time $T$, can be implemented precisely, while the other phase
gates definitely result in a decoherence in the qubit system, and can only
be implemented in an inaccurate way.

Obviously the Hamiltonian $H=H_{qc}+H_{c}$ describes a typical conditional
dynamics \cite{sun95}. Let the total system be initially in a superposition
of
\begin{equation}
|\Psi (0)\rangle =(c_{0}|0\rangle +c_{1}|1\rangle )\otimes |\alpha \rangle ,
\end{equation}
where the boson field is in a coherent state $|\alpha \rangle $. The total
system will evolve into an entangled state
\begin{equation}
|\Psi (t)\rangle =c_{0}|0\rangle \otimes |\alpha \rangle +c_{1}|1\rangle
\otimes e^{i\overset{\rightarrow }{\Phi }}|\alpha \rangle ,
\end{equation}
where
\begin{equation}
e^{i\overset{\rightarrow }{\Phi }}=\hat{T}\exp
(-i\int_{0}^{t}V_{I}(t^{\prime })dt^{\prime })
\end{equation}
is a time ordered integral. A formal phase operator can be explicitly
calculated as
\begin{equation}
\overset{\rightarrow }{\Phi }(t)=\eta (t)a+h.c+\varphi (t)+i\xi (t),
\end{equation}
where time-dependent coefficients
\begin{eqnarray}
\eta (t) &=&i\frac{g}{\omega }(1-e^{-i\omega t}),  \notag \\
\varphi (t) &=&\frac{|g|^{2}}{\omega ^{2}}(\omega t-\sin \omega t), \\
\xi (t) &=&\frac{|g|^{2}}{\omega ^{2}}(1-\cos \omega t),  \notag
\end{eqnarray}
are obtained through the Wei-Norman algebraic technique \cite{sun91}. Then
we can write down the total wave function as an entangled state%
\begin{equation}
|\Psi (t)\rangle =c_{0}|0\rangle \otimes |\alpha \rangle +e^{i\varphi
(t)-\xi (t)}c_{1}|1\rangle \otimes |\alpha +\eta (t)\rangle .
\end{equation}

Obviously, at the special instants
\begin{equation}
t=T=2k\pi /\omega ,
\end{equation}
where $k\in \mathsf{Z}$, both the decay factor $\xi (t)$ and the
displacement $\eta (t) $ in the coherent state $|\alpha +\eta (t)\rangle $
vanish. And a real phase
\begin{equation}
\varphi (T)=\phi _{s}=\frac{|g|^{2}}{\omega }T.
\end{equation}
occurs in the above entanglement state. Thus we realize a phase gate
operation Eq.(\ref{eqn:4}) of certain phase $\phi_{s}$, which is induced by
the factorized evolution
\begin{equation}
|\Psi (0)\rangle \rightarrow |\Psi (T)\rangle =(c_{0}|0\rangle
+c_{1}e^{i\phi _{s}}|1\rangle )\otimes |\alpha \rangle .
\end{equation}
It defines a reduced density matrix of a pure state
\begin{equation}
\rho _{q}=|c_{0}|^{2}|0\rangle \langle 0|+|c_{1}|^{2}|1\rangle \langle
1|+c_{1}c_{0}^{\ast }e^{i\phi _{s}}|1\rangle \langle 0|+h.c
\end{equation}
for the qubit system.

If the evolution time is not just at the instant $t=T$, the reduced density
matrix
\begin{equation}
\rho _{r}=|c_{0}|^{2}|0\rangle \langle 0|+|c_{1}|^{2}|1\rangle \langle
1|+c_{1}c_{0}^{\ast }D(t)|1\rangle \langle 0|+h.c
\end{equation}
is not of a pure state duo to the decoherence factor
\begin{equation}
D(t)=\langle \alpha |W(t)|\alpha \rangle =e^{i\phi (t)-\xi (t)}
\end{equation}%
where
\begin{equation}
\phi (t)=2\mathbf{Im}[\frac{(1-e^{-i\omega t})g\alpha }{\omega }]+\frac{%
(\omega t-\sin \omega t)|g|^{2}}{\omega ^{2}},
\end{equation}
and
\begin{equation*}
\xi (t)=\frac{|g|^{2}}{\omega ^{2}}(1-\cos \omega t).
\end{equation*}

The difference between $\rho _{q}$ and $\rho _{r}$ can be characterized by
the control fidelity $F(t)=Tr(\rho _{q}\rho _{r})$, which is defined as the
overlap of the target state $\rho _{q}$ and the final state $\rho _{r}$. By
a straightforward calculation, we have
\begin{eqnarray*}
F(t) &=&1-2|c_{0}|^{2}|c_{1}|^{2}[1-\mathbf{Re}(D(t)e^{i\phi _{s}})]  \notag
\\
&=&1-2|c_{0}|^{2}|c_{1}|^{2}[1-e^{-\xi (t)}\cos (\phi (t)-\phi _{s})].
\end{eqnarray*}
In Fig.1 we plot the curve $F(t)$, where $g=0.1$, $\omega =1$ and $\alpha
=1.5$. For convenience, we have take $|c_{0}|^{2}=|c_{1}|^{2}=1/2$ and then
\begin{equation}
F(t)=1-\frac{1}{2}[1-e^{-\xi (t)}\cos (\phi (t)-\phi _{s})].
\end{equation}
It can be seen that $F(t)$ is a periodic function with unity as the maximum
value. As a functional, the period $T=f[F]$ is a function of the function $%
F(t)$. $f[F]$ is determined by the system parameters $g$ and $\omega $. When
$t=f[F(t)]$, the control fidelity takes its maximum $F(t)=1$ and then we
realized an ideal phase gate operation with the phase $\phi
_{s}=|g|^{2}T/\omega $.

\begin{figure}[tp]
\centering
\includegraphics[scale=1]{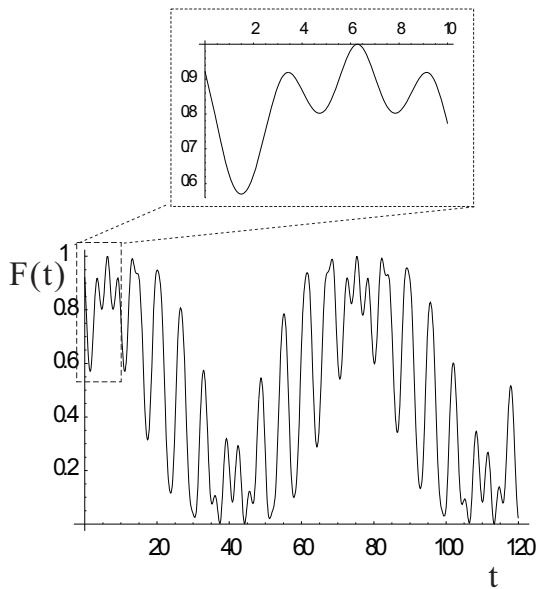}
\caption{The control fidelity $F(t)$ defined as the overlap of $\protect\rho%
_{q}$ and $\protect\rho _{r}$, which varies with $t$. The above inset is the
zoom of the curve, which indicates the collapses and revivals of the
controlled system. }
\label{fig1}
\end{figure}

In order to realize a real control we require that the effective interaction
$\langle V_{I}(t)\rangle $ could be automatically switched on and off at
time $0$ and $T$, i.e., the controllable condition ($CABC$)
\begin{equation}
\langle V_{I}(0)\rangle =\langle V_{I}(T)\rangle =0,
\end{equation}%
is satisfied for $\langle V_{I}(t)\rangle =\langle \psi
_{c}(t)|V_{I}(t)|\psi _{c}(t)\rangle $. For the above example, this
requirement means
\begin{eqnarray}
\mathbf{Re}(g\alpha ) &=&0,  \label{eqn109} \\
\mathbf{Im}[g\alpha ]\sin \omega T &=&-\omega \xi (T),  \label{eqn110}
\end{eqnarray}
for $\sin \omega T\neq 0$. When there is no loss of qubit coherence at the
instance $t=T$ ($\xi (T)=0$), the requirement Eq.(\ref{eqn109}) and Eq.(\ref%
{eqn110}) for an ideal quantum control is just $g\alpha =0$. It is absurd
and impracticable. However, there exist the situations ($\sin \omega T=0$)
satisfying the requirement for quantum control: $\mathbf{Re}[g\alpha ]=0$
and $\omega T=2k\pi $, $k\in \mathsf{Z}$, which is reasonable in principle
since a pure imaginary number $g\alpha =i|g\alpha |$ does not vanish even
though it have a vanishing real part. Therefore some target states are
obtained as the superpositions state of $|0\rangle $ and $|1\rangle $ with
specific relative phases that can be implemented perfectly by the quantum
control.

However, the above phase gate control could only generate particular phases $%
\phi _{s}$ on the qubit state $|1\rangle $, which is completely determined
by the coupling factor $g$ and the controller field frequency $\omega $. In
this sense we can not achieve a quantum control of implementing universal
phase gates for a given total system with fixed $g$ and the controller field
frequency $\omega $. To overcome this problem the local parameters of the
initial states of the controller should be used in the quantum control
rather than the fixed global parameters of the total system. We will explore
this possibility in Sec. \ref{sec:inequality} where the quantum decoherence
will be considered based on the uncertainty relation that relates to a
multi-mode coherent field.

\section{Quantum control by general controller}

\label{sec:general model}

Staring with an idealized model, the above investigations provide us some
insights into the quantum control problem. In order to consider the more
practical cases, we will analyze the quantum controllability in this
section. To focus on the central idea we do not consider the influence of
the environment yet. The entire system that we consider is an isolated
system including the controller $C$ with the Hamiltonian $H_{c}$ and the
controlled system $Q$ with the Hamiltonian $H_{q}$ . To bring out more
clearly the physical picture of such a quantum control, the minimal
assumption is that the Hamiltonian includes only two items: $H_{qc}$ and $%
H_{c}$. Matching this assumption, there exists a practical case that the
non-demolition control satisfies $[H_{q},H_{qc}]=0$ and then the free
evolution of the controlled system is eliminated.

Conveniently we work in the interaction picture with the Schr$\ddot{o}$%
dinger equation
\begin{equation}
i\hbar \frac{d}{dt}|\Psi ^{I}(t)\rangle =H_{qc}^{I}(t)|\Psi ^{I}(t)\rangle .
\label{eqn:eqn02}
\end{equation}%
Formally, the quantum control requires that the interaction Hamiltonian
\begin{equation}
H_{qc}^{I}(t)=e^{iH_{c}t/\hbar }H_{qc}e^{-iH_{c}t/\hbar }
\end{equation}%
can be automatically turned on and off at certain instants $t=0$ and $t=T$
during the evolution of the controller system. Under the quantum control a
quantum gate operation is accomplished by the controlled system. Besides, it
is also required that the controlling parameters depend on the initial state
of the controller system. By applying them to quantum computing, the quantum
computer implements the operations programmed by the controller.

Without loss of the generality, we still take the controlled system as a
qubit with two basis states $|0\rangle $ and $|1\rangle $. An ideal quantum
control with $U_{q}(T)$ exerting on the qubit can be described as a
factorized evolution
\begin{equation}
U_{I}(T)=e^{-\frac{i}{\hbar }(H_{qc}+H_{c})T}=U_{q}(T)\otimes U_{c}(T)
\end{equation}%
of the total system. So that a controlled evolution of the qubit system is
implemented as $|\psi _{q}(T)\rangle =U_{q}(T)|\psi _{q}(0)\rangle $, while $%
|\psi _{c}(T)\rangle =U_{c}(T)|\psi _{c}(0)\rangle $ defines the final state
of the controller. Here, $|\psi _{q}(0)\rangle =c_{0}|0\rangle
+c_{1}|1\rangle $ and $|\psi _{c}(0)\rangle $ are the initial states of the
qubit and the controller respectively. We note that, because the Hermitian
operators $H_{qc}$ and $H_{c}$ do not commute with each other, thus there is
not simply $\exp (-iH_{qc}T)=U_{q}(T)$ in practice. We emphasize that, due
to the limitation resulting from the Heisenberg uncertainty principle, the
realistic control can not be carried out in such a perfect way as a
completely factorized evolution.

Generally, the Hermitian operators $H_{qc}$ and $H_{c}$ do not commute with
each other and there exists an uncertainty relation:
\begin{equation}
\Delta H_{qc}^{I}\Delta H_{c}\geq \frac{1}{2}|\langle \lbrack
H_{qc}^{I},H_{c}]\rangle |,  \label{eqn:eqn4}
\end{equation}%
where the variations $\Delta A=\sqrt{\langle A^{2}-\langle A\rangle
^{2}\rangle }$, $A=H_{qc}$ and $H_{c}$. In a consistent approach for quantum
measurement \cite{zls}, this uncertainty relation is also responsible for
the decoherences induced by the detector as well as those induced by the
quantum control. Roughly speaking, the variation $\Delta H_{qc}^{I}(t)$ is
relavent to the induced decoherence in the qubit system, while the term $%
|\langle \lbrack H_{qc}^{I}(t),H_{c}]\rangle |$ indicates the influence of
quantum control, and $\Delta H_{c}$ is associated with the power or the
average energy of the controller. The conservation laws throw some limits on
such implementation of quantum gates \cite{Ozawa2002}. For example, a
quantum control to complete a CNOT gate usually concerns the transfer of
some conservation quantities between qubits. To focus on the problems in the
following, we will only consider the quantum control itself, which does not
involve the transfer of any known conservation quantities.

\begin{figure}[tp]
\begin{center}
\includegraphics[scale=0.70]{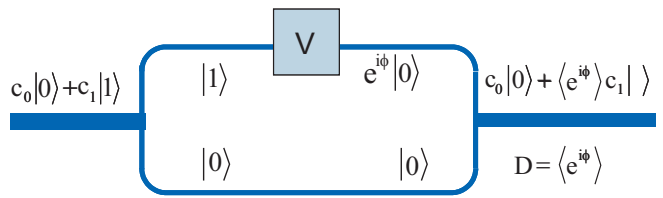}
\end{center}
\caption{(color online) Schematic illustration of the non-demolition
interaction for the phase quantum control: the effective potential $V$ only
act on the component $|1 \rangle$, but not on the component $|0 \rangle$ in
the superposition. Besides the wanted phase to be generated, such
interaction also induced a fluctuation of phase reflected by the factor $%
|D|\approx \exp(-\frac{1}{2}(\triangle \Phi)^2)$ with module less than 1.}
\end{figure}

Now we assume a non-demolition controlling interaction $H_{qc}=|1\rangle
\langle 1|\otimes V$ with a potential $V$ that acts only on the qubit state $%
|1\rangle $. It does not play any role at the beginning and the end of the
gate operation, but we require that it is generated by the controller, and a
nontrivial phase is left on the qubit state $|1\rangle $. Actually, as for
the quantum controls in quantum information processing, it is expected that
a quantum computer could work like electronic computers: when the programs
are designed then stored in it initially, the quantum computer should be
able to carry out computations without any other assistants. The basic
requirement for the quantum control is that the interaction can be switched
on and off automatically at certain instants, e.g., at $t=0$ and $t=T$,
\begin{eqnarray}
\langle V_{I}(0)\rangle &=&\langle \psi _{c}(0)|V_{I}(0)|\psi _{c}(0)\rangle
=0,  \notag \\
\langle V_{I}(T)\rangle &=&\langle \psi _{c}(T)|V_{I}(T)|\psi _{c}(T)\rangle
=0,  \label{eqn:c-condition}
\end{eqnarray}%
where $V_{I}(t)\equiv \exp (iH_{c}t/\hbar )V(-iH_{c}t/\hbar )$. The above
Eq.(\ref{eqn:c-condition})is the general controllable condition. The
sandwich $\langle V\rangle $ is defined as the average of the operator $V$
over the controller state. This means that the effective interaction is
obtained by taking the average of $V_{I}(t)$ over the instantaneous
controller states $|\psi _{c}(t)\rangle $.

Generally, the controller in physical implementations of the quantum control
are various fields that are supposed to be classical. For example, the
microwave electromagnetic fields are used to manipulate the nuclear
spin-qubits in NMR, the laser fields are applied to control the atomic
qubits and the classical magnetic flux and voltage are utilized to adjust
the Josephson-Junction based qubits. However, the controlling fields are
essentially of quantization and are usually described by coherent states or
some quantum mechanical mixture.

Starting from the initial state where the qubit is in $|\psi _{q}\rangle
=c_{0}|0\rangle +c_{1}|1\rangle $, the total system evolves according to the
entangled state
\begin{equation}
|\Psi (t)=c_{0}|0\rangle \otimes |\psi _{c}(0)\rangle +c_{1}|1\rangle
\otimes e^{i\mathbf{\Phi }(t)}|\psi _{c}(0)\rangle ,  \label{eqn:eqn5}
\end{equation}%
where we have defined the time-order integral
\begin{equation}
e^{i\mathbf{\Phi }}=\hat{T}\exp (-\frac{i}{\hbar }\int_{0}^{t}V_{I}(\tau
)d\tau ).  \label{eqn:33}
\end{equation}%
The decoherence factor \cite{sun93} is an expectation of the unitary
operator
\begin{equation}
D(T)=\langle \psi _{c}(0)|e^{i\mathbf{\Phi }}|\psi _{c}(0)\rangle ,
\end{equation}
which can be used to characterize the quantum controllability.

Now we need to consider that in what cases the above entangled state $|\Psi
(t)\rangle $ can become a factorized state Eq.(\ref{eqn:eqn3}) at ceratain
instant $t=$ $T$ so that the ideal quantum control is realized by choosing
the initial state $|\psi _{c}(0)\rangle $ of the controlling system. A
simplest illustration is that $V_{I}(t)=V$ is a static potential and thus
\begin{equation}
e^{i\mathbf{\Phi }}=\exp (-iTV/\hbar ).
\end{equation}%
If we choose $|\psi _{c}(0)\rangle =|\phi \rangle $ with the eigenvalue $%
\phi $, then $\exp (i\mathbf{\Phi })$ becomes a $c-number$ phase factor $%
\varphi $, and the time evolution automatically generate a phase gate
operation with the $c-number$ phase:
\begin{equation}
|\Psi (T)=(c_{0}|0\rangle +c_{1}e^{i\varphi }|1\rangle )\otimes |\psi
_{c}(0)\rangle .
\end{equation}%
Indeed, the phase $\phi $ multiplied to the qubit state $|1\rangle $ is well
defined and can be generated with arbitrary precision at a suitable instant $%
T$ by choosing the initial state $|\psi _{c}(0)\rangle =|\phi \rangle $ of
the controller. This is what we want: the qubit system is controlled by the
parameters of the initial state as well as the evolution time. It seems that
no fundamental restrictions exists for $|\psi _{c}(0)\rangle $ and $T$.

However, the above idealized situation is far from the realistic cases in
practical quantum controls. Firstly, the precision of the quantum control is
guaranteed by the stability of potential, i.e., $[V,H_{c}]=0=[H_{qc},H_{c}]$%
. However, this means that the free Hamiltonian evolution of controller has
no influence on the effective interaction by $H_{qc}$ and thus the $CABC$
can not be satisfied automatically. Therefore, we infer that, in order to
realize a quantum control with the ``switched on and off", the potential $%
V_{I}(t)$ could not be a static one. In this case the $c-number$ phase is
not well defined by the initial state of the controller and thus there
exists a phase fluctuation $\Delta \Phi$ in the implementation of the
quantum control.

To explore the possibility of assorting with the $CABC$ and the precision of
the quantum control, we distinguish two cases by whether the potential $%
V_{I}(t)$ generated by the controller is commutative or not at different
instants, i.e.,
\begin{eqnarray}
Case.1 &:&[V_{I}(t),V_{I}(t^{\prime })]=0,  \label{eqn:case-a} \\
Case.2 &:&[V_{I}(t),V_{I}(t^{\prime })]\neq 0.  \label{eqn:case-b}
\end{eqnarray}

In the first case a phase factor operator can be simply defined as
\begin{equation}
\mathbf{\Phi }=-\frac{1}{\hbar }\int_{0}^{T}V_{I}(\tau )d\tau .
\end{equation}%
Under the small variation $\Delta \mathbf{\Phi }\ll 1$, the decoherence
factor can be calculated as
\begin{equation}
D(T)\approx e^{i\langle \mathbf{\Phi }\rangle -\frac{1}{2}(\Delta \mathbf{%
\Phi })^{2}}\equiv e^{i\langle \mathbf{\Phi }\rangle }d(T).
\end{equation}%
Similar to the arguments about the exactly solvable model in Sec.\ref%
{sec:single}, an observation is that the ideal quantum control can be
characterized by whether or not the decoherence factor $| D(T)| =| \langle
\exp (i \mathbf{\Phi )}\rangle | $ can reach unity. Actually the phase
multiplied to the qubit state $|1\rangle $ is the real part of the
expectation value of the phase factor operator $\langle \mathbf{\Phi }
\rangle $ plus a decay factor from its quantum fluctuation $(\Delta \mathbf{%
\Phi )}^{2}$\cite{Aha}.Thus the quantum controllability is destroyed by the
phase fluctuation $(\Delta \mathbf{\Phi )}^{2}$ in general.

In the following, we will show that the phase fluctuation $(\Delta \mathbf{%
\Phi )}^{2}$ will result in a loss of quantum coherence or quantum
dephasing. To this end we calculate
\begin{equation}
(\Delta \mathbf{\Phi })^{2}\sim \frac{1}{\hbar ^{2}}\int_{0}^{T}dt%
\int_{0}^{t}\Delta V_{I}(t)\Delta V_{I}(\tau )d\tau ,
\end{equation}%
which shows that phase fluctuation $(\Delta \mathbf{\Phi )}^{2}$ is just the
correlated fluctuation of Heisenberg interaction. Thus $d(T)=\exp [-(\Delta
\mathbf{\Phi })^{2}/2]$ is a decaying factor in $D(T)$ accompanying the
off-diagonal terms of the reduced density matrix of the qubit system. To
quantitatively describe that to what extent the target sate
\begin{equation*}
|\psi _{t}\rangle =c_{0}|0\rangle +c_{1}e^{i\langle \mathbf{\Phi }\rangle
}|1\rangle
\end{equation*}%
can be reached by the controlled time evolution $|\Psi (t)\rangle$, the
control fidelity
\begin{eqnarray}
F(t) &=&Tr[|\Psi (t)\rangle \langle \Psi (t)|(1\otimes |\psi _{t}\rangle
\langle \psi _{t}|)]  \notag \\
&=&Tr_{c}(\langle \psi _{t}|\Psi (t)\rangle \langle \Psi (t)|\psi
_{t}\rangle )=Tr(\rho _{t}\rho _{r})
\end{eqnarray}%
is defined in terms of the reduced density matrix $\rho _{t}$ and the
reduced density matrix $\rho _{r}=Tr_{c}[|\Psi (t)\rangle \langle \Psi (t)|]$%
, where $Tr_{c}$ indicates tracing over the variables of the controller. In
this case the result is obtained as
\begin{equation*}
F(t)=1-2|c_{0}|^{2}|c_{1}|^{2}(1-e^{-\frac{1}{2}(\Delta \mathbf{\Phi }%
)^{2}}).
\end{equation*}%
Thus the corresponding error measure
\begin{equation}
\varepsilon =1-F(t)=2|c_{0}|^{2}|c_{1}|^{2}[1-d(t)]
\end{equation}%
describes the failure probability of the quantum control.

For the second case, due to the non-vanishing commutator between $V_{I}(t)$
at different instants we can not generally define a phase factor operator $%
\mathbf{\Phi }$, but we can still formally write $D(T)=\langle \exp (i%
\mathbf{\Phi })\rangle $ or
\begin{equation}
D(T)=e^{i\Phi-\xi }\simeq \exp (i\langle \mathbf{\Phi }\rangle -\frac{1}{2}%
(\Delta \mathbf{\Phi })^{2}).
\end{equation}%
This can give all similar results as the case 1. The exactly solvable model
in Sec. \ref{sec:single} belongs to the second case. This result is exact
for the above example presented in the last section where
\begin{equation}
\frac{1}{2}(\Delta \Phi )^{2}=\xi (t),\langle \mathbf{\Phi }\rangle =\phi
(t).
\end{equation}

As discussed in the above, the decoherence induced limit to the quantum
control has been explained based on the phase uncertainty. In fact, this
understanding reveals once again the inherence of the quantum decoherence in
the generalized two-slit experiment about $|0\rangle $ and $|1\rangle $,
whose interference fringe vanishes when one determines which slit the
particle comes from. According to Heisenberg, this is due to the randomness
of relative phases \cite{heisenberg} from the quantum control. Furthermore,
we can conclude from the above exact solution that the large random phase
change just originates from Heisenberg's position-momentum uncertainty
relation $\Delta x_{k}\Delta p_{k}=1/2$. This observation will help us to
discover a bound on the quantum control.

\section{Phase uncertainty due to standard quantum limit}

\label{sec:uphase}

Based on our previous explorations on the relation between the two
explanations for quantum decoherence \cite{zls}, using the position-momentum
uncertainty relation, we now can associate the physical limit of quantum
control with the standard quantum limit (SQL) in quantum measurement context
\cite{SQL} through a concrete example as follows.

This is a more practical example that the qubit is controlled by a
multi-mode electromagnetic field
\begin{equation}
E=\sum_{k}(u_{k}(x)a_{k}e^{-i\omega _{k}t}+h.c)
\end{equation}%
with the mode functions $u_{k}(x)$. The controlling Hamiltonian $%
H_{qc}(t)=|1\rangle \langle 1|\otimes V_{I}(t)$ in the interaction picture
reads as
\begin{eqnarray}
H_{qc}(t) &=&|1\rangle \langle 1|\otimes \sum_{k}H_{k}  \notag \\
&\equiv &|1\rangle \langle 1|\otimes \sum_{k}\hbar (g_{k}a_{k}e^{-i\omega
_{k}t}+h.c),
\end{eqnarray}%
where $\omega _{k}$ are the mode frequencies, $a_{k}$ and $a_{k}^{\dag }$
the creation and annihilation operators respectively, and $g_{k}$ the mode
couplings constants between the qubit and the field modes. We suppose that
the electromagnetic field is initially prepared in a multi-mode coherent
state
\begin{equation}
|\psi _{c}(0)\rangle =|\mathbf{\alpha }\rangle \equiv \prod_{k}|\alpha
_{k}\rangle
\end{equation}%
as a direct product of the coherent state $|\alpha _{k}\rangle $ of $k$th
mode. In such a initial state, the observable is the the average of the
field operator
\begin{equation}
\langle \alpha |E|\alpha \rangle =\sum_{k}[u_{k}(x)\alpha _{k}e^{-i\omega
_{k}t}+h.c],
\end{equation}%
which is a wave packet, the superposition of many plane waves. This means
that, to realize a more realistic quantum control, we need a wave packet
rather than a single mode or a plane wave.

The free Hamiltonian of the qubit system has been omitted without loss of
generality. The potential $V_{I}(t)$ exerts on the qubit state $|1\rangle ,$
but not on the qubit state $|0\rangle $. Then the evolution can be obtained
as
\begin{equation*}
U(t)=|0\rangle \langle 0|\otimes \mathbf{1+}|1\rangle \langle 1|\otimes e^{i%
\mathbf{\Phi }},
\end{equation*}%
where
\begin{equation}
e^{i\mathbf{\Phi }}=\prod\limits_{k}e^{i\Phi
_{k}}=\prod\limits_{k}U_{k}\equiv \prod\limits_{k}\hat{T}\exp (-\frac{i}{%
\hbar }\int_{0}^{t}H_{k}(\tau )d\tau ).
\end{equation}%
We can explicitly calculate the phase operator $\Phi =\sum_{k}\Phi _{k}$
defined above by the method similarly to that used for the example about the
single mode field in Sec. \ref{sec:single}. It is obtained by
\begin{equation*}
\Phi _{k}=\eta _{k}(t)a_{k}+h.c+\varphi _{k}(t)+i\xi _{k}(t),
\end{equation*}%
with three time-dependent parameters
\begin{eqnarray}
\eta _{k}(t) &=&i\frac{g_{k}}{\omega _{k}}(1-e^{-i\omega _{k}t}),  \notag \\
\varphi _{k}(t) &=&\frac{|g_{k}|^{2}}{\omega _{k}^{2}}(\omega _{k}t-\sin
\omega _{k}t), \\
\xi _{k}(t) &=&\frac{|g_{k}|^{2}}{\omega _{k}^{2}}(1-\cos \omega _{k}t).
\notag
\end{eqnarray}%
The phase operator can be re-written as $\mathbf{\Phi }=\Omega (t)+\mathbf{%
\Phi }_{a}$ in terms of the constant phase $\Omega (t)=\sum_{k}\varphi
_{k}(t)$ plus the operator
\begin{equation}
\mathbf{\Phi }_{a}=\sum_{k}\mathbf{\Phi }_{ak}(t)=\sum_{k}(\eta
_{k}(t)a_{k}+h.c).
\end{equation}

The decoherence factor can be calculated similarly as
\begin{equation}
D(t)=\langle \mathbf{\alpha }|e^{i\mathbf{\Phi }}|\mathbf{\alpha }\rangle
=e^{i\phi (t)-\xi (t)},
\end{equation}%
where $\xi (t)=\sum_{k}\xi _{k}(t)$ and
\begin{eqnarray}
\phi (t) &=&2\sum_{k}\{\mathbf{Im}[\frac{g_{k}\alpha _{k}}{\omega _{k}}%
(1-e^{-i\omega _{k}t})]  \notag \\
&&+\sum_{k}\frac{|g_{k}|^{2}}{\omega _{k}^{2}}(\omega _{k}t-\sin \omega
_{k}t).
\end{eqnarray}%
It is easy to check that the phase generated by the quantum control is just
the average value of the phase operator
\begin{eqnarray}
\langle \mathbf{\alpha }|\mathbf{\Phi }|\mathbf{\alpha }\rangle &=&\langle
\alpha |\mathbf{\Phi }_{a}|\alpha \rangle +\Omega (t)  \notag \\
&=&\sum_{k}(\eta _{k}(t)\alpha _{k}+h.c)+\sum_{k}\varphi (t).
\end{eqnarray}%
The analytical expression of the phase fluctuation is
\begin{eqnarray}
(\Delta \Phi )^{2} &=&(\Delta \Phi _{a})^{2}=\sum_{k=1}^{N}(\Delta \Phi
_{ak})^{2} \\
&=&\sum_{k=1}^{N}|\eta _{k}|^{2}=2\sum_{k=1}^{N}\xi _{k}(t)=2\xi (t),  \notag
\end{eqnarray}%
where we have considered each uncertain phase change as an independent
stochastic variable. Namely, the relation $\xi (t)=(\Delta \Phi )^{2}/2$ or
the exact expression $D(T)=\exp (i\langle \mathbf{\Phi }\rangle -(\Delta
\mathbf{\Phi })^{2}/2)$\ still holds for the multi-mode case with the
specialized initial state. Correspondingly, the error measure is estimated
as
\begin{equation}
\varepsilon =1-F(t)=\lambda (\Delta \mathbf{\Phi }_{a})^{2}=2\lambda \xi (t),
\end{equation}
where $\lambda =|c_{0}|^{2}|c_{1}|^{2}$. Different from the single mode
case, it is hard to find a proper instant $T$ such that $\varepsilon =\xi
(T)=0$ in general. Namely, it is hard to achieve an ideal quantum control
without any error.

In the above discussions, the realization of quantum control boils down to
the appearance of the $c$-number phase $\phi (t)$ that contains the
controllable part depending on the initial state of the controller. An ideal
quantum control means the vanishing error $\lambda (\Delta \mathbf{\Phi }%
_{a})^{2}$. But it is almost impossible because of the intrinsic decoherence
due to quantum control itself. In fact, if the electromagnetic field could
carry out a completely efficient control on the controlled system, then the
interaction Hamiltonian should not commute with that of the controller.
These facts are responsible for the inaccuracy of the phase gate or
decoherence in the controlled system under the quantum control. We have to
point out that the conclusion drawn above seems to depend on the choice of
the initial state, but now we can argue that this is not the case with the
above consideration. So we need to consider the universality of the
conclusions.

Physically, every variable of the controller can independently exert a
different impact on the different components of controller state. Since
every uncertain phase is an independent stochastic variable, we have
\begin{equation}
(\Delta \mathbf{\Phi _{a}})^{2}=\sum_{k=1}^{N}(\Delta \mathbf{\Phi }%
_{ak})^{2}\geq N\min \{(\Delta \mathbf{\Phi }_{ak})^{2}|k=1,2..N\}  \notag
\end{equation}%
for a general initial state of the controller. We note that the phase
uncertainty $(\Delta \mathbf{\Phi _{a}})^{2}$ caused by the controller
variables can be amplified to a number much larger than unity when $N
\rightarrow \infty $, i.e., the system states acquire a very large random
phase factor. The decay factor
\begin{equation*}
|D(t)|=e^{-\frac{1}{2}(\Delta \mathbf{\Phi })^{2}}\leq \exp (-\frac{N}{2}%
\min \{(\Delta \mathbf{\Phi }_{ak})^{2}|).
\end{equation*}%
So $|D(t)|\rightarrow 0$ when $N\rightarrow \infty $, i.e., the macroscopic
controller can wash out the quantum coherence of the controlled system.

To be more concrete we assume that, in the initial state $|0\rangle =\otimes
_{k=1}^{N}|\psi _{k}\rangle $ of the controller, each component $|\psi
_{k}\rangle $ is a wave packet, symmetric with respect to both the
\textquotedblleft canonical coordinate" $x_{k}=(a_{k}+a_{k}^{\dagger })/%
\sqrt{2}$ and the corresponding \textquotedblleft canonical momentum" $%
p_{k}=-i(a_{k}-a_{k}^{\dagger })/\sqrt{2}$. So $\langle x_{k}\rangle \equiv
\langle \psi _{k}|x_{k}|\psi _{k}\rangle =0$ and $\langle p_{k}\rangle =0$.
We do not need the concrete form of the initial state. For convience we
assume it to be of Gaussian type with the variance $\sigma _{k}$ $=\Delta
x_{k}$ in $x_{k}-space$. Physically, once $\Delta x_{k}$ is given, the
variance of $p_{k}$ cannot be arbitrary since there is a Heisenberg's
position-momentum uncertainty relation $\Delta x_{k}\Delta p_{k}\geq 1/2$.
In the following we will show that the uncertainty relation will give a low
bound to the variance of $\Delta \mathbf{\Phi }_{a}$. In the above reasoning
about $|D(t)|\rightarrow 0$ when $N\rightarrow \infty $, we have considered
that there exists a finite minimum value of $(\Delta \mathbf{\Phi }%
_{ak})^{2} $. In the quantum measurement theory, the finite minimum value of
$(\Delta \mathbf{\Phi }_{ak})^{2}$ is implied by the so called \ standard
quantum limit (SQL) on the continuous measurement of phase operator.

To see this we rewrite \ the phase operator
\begin{equation}
\mathbf{\Phi }_{a}=\sum_{k}\mathbf{\Phi }_{ak}=\sum_{k}[\varkappa
_{k}(t)x_{k}+\mu _{k}(t)p_{k}],
\end{equation}%
in terms of the \textquotedblleft canonical coordinate" and the
corresponding \textquotedblleft canonical momentum", and the coefficients
are
\begin{eqnarray*}
\varkappa _{k}(t) &=&\frac{1}{\sqrt{2}}(\eta _{k}(t)+\eta _{k}^{\ast }(t)),
\\
\mu _{k}(t) &=&\frac{i}{\sqrt{2}}(\eta _{k}(t)-\eta _{k}^{\ast }(t)).
\end{eqnarray*}%
The existence of SQL is guaranteed by the Heisenberg's position-momentum
uncertainty relation$.$ Because each $\mathbf{\Phi }_{ak}\ =\varkappa
_{k}(t)x_{k}+\mu _{k}(t)p_{k}$ is a linear combination of $x_{k}$ and $p_{k}$
with a property $\langle x_{k}p_{k}\rangle +\langle p_{k}x_{k}\rangle =0$
for the average over the real initial state. The phase fluctuation $\Delta
\mathbf{\Phi }_{ak}$ can be derived as
\begin{eqnarray*}
\Delta \mathbf{\Phi }_{ak} &=&\sqrt{|\varkappa _{k}(t)|^{2}(\Delta
x_{k})^{2}+|\mu _{k}(t)|^{2}(\Delta p_{k})^{2}} \\
&\geq &\sqrt{|\varkappa _{k}(t)\mu _{k}(t)|},
\end{eqnarray*}%
or
\begin{equation}
(\Delta \mathbf{\Phi }_{ak})^{2}\geq 8\frac{g_{k}^{2}}{\omega _{k}^{2}}|\sin
^{3}\frac{\omega _{k}t}{2}\cos \frac{\omega _{k}t}{2}|.
\end{equation}%
Here, we considere the variance $\Delta (\xi x)=|\xi |(\Delta x)$ for a
stochastic variable $x$ and a real number $\xi $, and suppose $g_{k}/\omega
_{k}$ being a real number.

In the above arguments, $x_{k}$ and $p_{k}$ are not only regarded as a pair
of uncorrelated stochastic variables in the terminology of classical
stochastic process, the uncertainty relation $\Delta x_{k}\Delta p_{k} \sim
1/2$ of them is also taken into account. This constraint just reflects the
uncertainty of phase change in the quantum control process. Therefore, we
have a time dependent minimum value of phase uncertainty with a low bound
\begin{equation}
(\Delta \mathbf{\Phi _{a}})^{2}\geq N\min \{|\varkappa _{k}(t)\mu
_{k}(t)||k=1,2..N\}.  \notag
\end{equation}%
This result qualitatively illustrates the many-particle amplification effect
of uncertain phase change due to quantum control itself. The large random
phase variance $(\Delta \mathbf{\Phi _{a}})^{2}$ implies that it is hard to
satisfy the exact condition $(\Delta \mathbf{\Phi _{a}})^{2}=0$ in
principle, and thus one can only optimize both the system parameters and the
initial state of the controller to approach what we want.

To see the above observation analytically, we calculate $\langle \mathbf{%
\Phi }\rangle $ in comparison with $\Delta \mathbf{\Phi }$ in the
decoherence factor $D(T)=\exp (i\langle \mathbf{\Phi }\rangle -(\Delta
\mathbf{\Phi })^{2}/2)$. The most simple, but somewhat trivial case is that
all modes are degenerate, i.e., $g_{k}=g$ and $\omega _{k}=\omega$, then
\begin{equation}
\Delta \mathbf{\Phi }=\sqrt{8N}\frac{|g|}{\omega }|\sin \frac{\omega t}{2}|,
\end{equation}
while the phase we wanted is
\begin{eqnarray}
\phi (t) &=&2N\{\mathbf{Im}[\frac{g\alpha }{\omega }(1-e^{-i\omega t})]
\notag \\
&&+2N\frac{|g|^{2}}{\omega ^{2}}(\omega t-\sin \omega t).
\end{eqnarray}
Obviously, for very large $N$, the phase fluctuation $\Delta \mathbf{\Phi }$
can be neglected since $\Delta \mathbf{\Phi /|}\phi (t)|\sim 1/\sqrt{N}
\rightarrow 0$. In general, we need to consider divergence of the phase
fluctuation
\begin{eqnarray}
(\Delta \mathbf{\Phi })^{2} &=&\sum_{k=1}^{N}16\frac{g_{k}^{2}}{\omega
_{k}^{2}}\sin ^{2}\frac{\omega _{k}t}{2}  \notag \\
&=&\int_{-\infty }^{\infty }16\frac{g_{k}^{2}}{\omega _{k}^{2}}\rho (\omega
_{k})\sin ^{2}\frac{\omega _{k}t}{2}d\omega _{k}
\end{eqnarray}
for various spectrum distributions of the controller, where an unspecific
spectrum distribution $\rho (\omega _{k})$ is used to discuss the case with
continuous spectrum. For example, when $\rho (\omega _{k})=\gamma /g_{k}^{2}$%
, the decoherence factor is exponentially decaying since the above integral
converges to a number $8\pi \gamma t/9$ proportional to time $t$. Another
example is the Ohmic distribution $\rho (\omega _{k})=2\eta \omega
_{k}^{2}/(\pi g_{k}^{2})$, which results in a diverging phase fluctuation
for $t\neq 0$.

\section{Low bound of the control induced decoherence and quantum computation%
}

\label{sec:inequality}

In this section we will show that, it is the back-action of the controller
on the controlled system, implied by Heisenberg's position-momentum
uncertainty relation, that disturb the phases of states of the controlled
system and then induce a quantum decoherence, which is relevant to the SQL.
In order to quantitatively characterized such limit to the quantum
controllability, we now return to the discussion about the quantum control
with multi-mode field initially prepared in a coherent state.

The commutation relation of the number operator $\mathbf{N}$ and
the phase operator $\mathbf{\Phi }_{a}$\ defines an operator
\begin{equation}
\mathbf{\Theta }=i\sum_{k}(-\eta _{k}(t)a_{k}+\eta _{k}^{\ast
}(t)a_{k}^{\dag })
\end{equation}%
dual to the phase operator $\mathbf{\Phi }_{a},$that is
\begin{equation}
\mathbf{\Theta }=i[\mathbf{N},\mathbf{\Phi }_{a}].
\end{equation}%

To see the meaning of the defined $\mathbf{\Theta ,}$ we calculate
the commutation relation of $\mathbf{N}$ and $\mathbf{\Phi }_{a}$
to find a close algebra by
\begin{eqnarray}
\lbrack \mathbf{N},\mathbf{\ \Theta }] &=&i\mathbf{\Phi }_{a},  \notag \\
\lbrack \mathbf{\Phi }_{a},\mathbf{\Theta }] &=&iF(t)
\end{eqnarray}%
where $F(t)=2\sum_{k}|\eta _{k}(t)|^{2}$ is a time dependent constant. This
means that $\ \mathbf{P=\Theta /}F(t)$ is a conjugate variable with respect
to $\mathbf{\Phi }_{a}$ since we have the canonical commutation relation $[%
\mathbf{\Phi }_{a},\mathbf{P}]=1.$  In this sense we call
$\mathbf{\Theta }$ a dual phase operator (DPO).  A constant
uncertainty relation can be found for $\mathbf{\Phi }_{a}$ and
$\mathbf{P}$ , \ which can be minimized by \ the corresponding
coherent state $|\alpha \rangle =\prod_{k}|\alpha _{k}\rangle $.

The above arguments  about minimization of the uncertainty by  $[\mathbf{%
\Phi }_{a},\mathbf{\Theta }]$ can enlighten us to find a low bound
for the control induced decoherence. To this end  we consider the
uncertainty
relation%
\begin{equation}
\langle \mathbf{N}\rangle \Delta \mathbf{\Phi }_{a}=\Delta \mathbf{N}\Delta
\mathbf{\Theta }\geq \frac{1}{2}|\langle \lbrack \mathbf{\Theta },\mathbf{N}%
]\rangle |=\frac{1}{2}|\langle \mathbf{\Phi }_{a}\rangle |  \label{eqn:aa}
\end{equation}%
about DPO and the photon number operator
$\mathbf{N}=\sum_{k}a_{k}^{\dag }a_{k}$.

To derive the above uncertainty relation (\ref{eqn:aa}), we have
considered
\begin{eqnarray}
\Delta \mathbf{N} &=&\langle \mathbf{N}\rangle ,  \notag \\
(\Delta \mathbf{\Phi }_{a})^{2} &=&(\Delta \mathbf{\Theta )}^{2}
\label{eqn:xxx}
\end{eqnarray}%
for the average $\langle \mathbf{\ }\rangle $ over the coherent state $%
|\alpha \rangle  $. We check the above results (%
\ref{eqn:xxx}) by the straightforward calculations
\begin{eqnarray}
(\Delta \Phi _{a})^{2} &=&\langle \alpha |\Phi _{a}^{2}-\langle \Phi
_{a}\rangle ^{2}|\alpha \rangle =\sum_{k}|\eta _{k}(t)|^{2},  \notag \\
(\Delta \mathbf{\Theta )}^{2} &=&\langle \alpha |\mathbf{\Theta }%
^{2}-\langle \mathbf{\Theta }\rangle ^{2}|\alpha \rangle =\sum_{k}|\eta
_{k}(t)|^{2}.  \notag
\end{eqnarray}

The novel uncertainty relation (\ref{eqn:aa}) defines a low bound for the
phase variation $\Delta \Phi _{a}$ for a given phase $\langle \mathbf{\Phi }%
_{a}\rangle $ to be achieved by the quantum control, i.e.
\begin{equation}
\Delta \mathbf{\Phi }_{a}\geq \frac{|\langle \mathbf{\Phi }_{a}\rangle |}{%
2\langle \mathbf{N}\rangle }.  \label{eqn:maine}
\end{equation}%
Eq.(\ref{eqn:maine}) clearly implies that we need much larger energy to
reduce the low bound of the phase fluctuation. Actually, we can formally
write down the expectation of the photon energy of the controller
\begin{equation}
E=\langle {\hbar }\sum_{k}\omega _{k}a_{k}^{\dag }a_{k}\rangle \equiv {\hbar
}\langle \mathbf{N}\rangle {\langle \omega \rangle }
\end{equation}%
in terms of the average photon number $\langle \mathbf{N}\rangle
=\sum_{k}|\alpha _{k}|^{2}$ and the average frequency of photons
\begin{equation}
{\langle \omega \rangle }{=}\frac{\sum_{k}\omega _{k}|\alpha _{k}|^{2}}{%
\sum_{k}|\alpha _{k}|^{2}}.
\end{equation}%
Then Eq.(\ref{eqn:maine}) becomes
\begin{equation}
\Delta \Phi _{a}\geq \frac{\hbar \langle \omega \rangle }{2E}|\langle
\mathbf{\Phi }_{a}\rangle |.
\end{equation}%
The small low bound requires that a large quantum controller
(implied by large $\langle \mathbf{N}\rangle $ or large energy
$E$) \ possesses a very small average frequency. In this sense
Eq.(\ref{eqn:aa}) defines a necessary condition for the quantum
control that can manipulate the qubit system reaching the target
state accurately. This requirement is very similar to that the
apparatus should be sufficiently \ so \textquotedblleft large"
that to be "classical" in the quantum measurement in the so-called
\textquotedblleft Copenhagen interpretation". Since the quantum
control relies on the ability to preserve quantum coherence of the
qubit system during controlling it, the controller should be much
\textquotedblleft larger" than the controlled system. In this
sense, the bac-action of the qubit system on the controller can be
neglected.

Next we consider the controllable condition (\ref{eqn:c-condition}) that the
controller field is switched on and off over a time duration $T$, which can
be roughly realized as a periodical phenomenon with the average period $%
T\sim 2\pi /\langle \omega \rangle $. Since the average frequency of the
field can be approximated by $\langle \omega \rangle \sim 2\pi /T$, there is
a low bound
\begin{equation}
\varepsilon \geq \frac{\lambda h^{2}}{4E^{2}T^{2}}|\langle \mathbf{\Phi }%
_{a}\rangle |^{2}\equiv \frac{\lambda h^{2}}{4S^{2}}|\langle \mathbf{\Phi }%
_{a}\rangle |^{2}  \label{eqn:inequality}
\end{equation}%
for the error measure estimation of $\varepsilon \sim \lambda (\Delta
\mathbf{\Phi }_{a})^{2}$ of the quantum control. \ So the larger action $%
S=ET $ from the controller is brought on the qubit system, the less quantum
decoherence characterized by the control induced error $\varepsilon $
becomes; the more one wants to change by the phase $\langle \mathbf{\Phi }%
_{a}\rangle $ of the qubit system, the larger quantum decoherence is induced
by the quantum control. Therefore Eq.(\ref{eqn:inequality}) imposes a
fundamental limit on the accuracy of quantum control. In the following we
can consider this physical limit for quantum computing

\begin{figure}[tp]
\begin{center}
\includegraphics[scale=0.8]{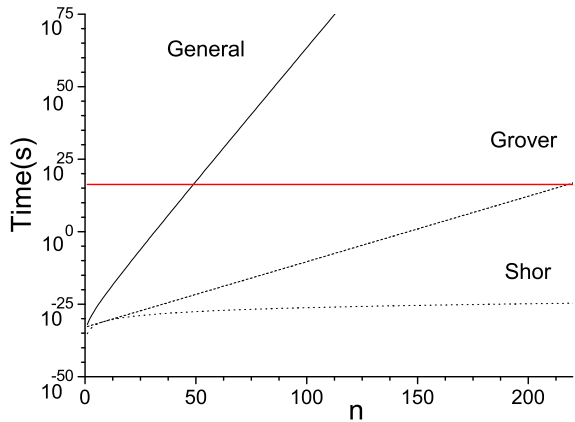}
\end{center}
\caption{The minimum amount of time needed for some algorithms, in
logarithmic scale. $n$ is the number of qubits consisted in the quantum
algorithm. The horizontal line at about $10^{16}$ indicates the age of
universe.}
\label{algorithms}
\end{figure}

%complexity(L):
%General: n^24^n;
%Grover: n^(1/2);
%Shor: (log2)^3n^3.
%age of the universe: 150*10^8 year= 1.97*10^16 second.
%T_total=\frac{\pi^2 h}{4\sqrt{2}100}*L^2=1.2*10^(-33)*L^2

%T_total= \frac{h \lambda L^2 \pi}{2 E }
%T_total= \frac{h}{ E }L^3/2

It is well known that the controllability of qubits is a basic requirement
for universal quantum computations, but according to the above arguments a
well-armed controls in quantum computing would cause the extra decoherence
in the qubits system. Thus in competition with the induced decoherence, the
controllability for quantum computation is limited.

In the last section a low bound of decoherence from the quantum control is
obtained. It throws an accuracy limit to the quantum controls in quantum
computation. According to Eq.(\ref{eqn:inequality}) this limit is about $%
10^{-20}$ for the typical setting $\langle \Phi _{a}\rangle =\pi $, $%
E=10^{-9}J$ and $T=1\mu s$ in an ion trap schemes. This is such a small
limit that it is negligible in comparison with other errors, such as the
environments induced errors in the current experiments of implementing
quantum computation. However, in principle, Eq.(\ref{eqn:inequality}) do
throw a fundamental limit to the accuracy of the quantum control and thus on
quantum computations. There are some numerical estimates in Fig.\ref%
{algorithms}, which demonstrate the similar limit to the power of quantum
computer. It is known that for an algorithm consisting of $L$ operations on
the qubits system, an up bound of error $\epsilon \sim 1/L$ is required in
each operation for a faithful result of the entire computation. The
inequality Eq.(\ref{eqn:inequality}) tells us that the minimum amount of
time needed for a single gate is
\begin{equation}
t_{\min }=\frac{h\sqrt{\lambda L}}{2E}|\langle \Phi _{a}\rangle |
\end{equation}%
and so the total time needed to carry out a particular algorithm consisting
of $L$ elementary gates is about $Lt_{\min }$.

For a general algorithm as an arbitrary unitary operation on $n-qubit$, the
amount of elementary gates $L$ needed is about $\mathit{O}(n^{2}4^{n})$ \cite%
{Barenco}; for the Grover algorithm on $n-qubit$, the amount is about $%
\mathit{O}(\sqrt{2^{n}})$; for the Shor large number factorization, the
amount is about $\mathit{O}(n^{3}log^{3}2)$. The time duration needed for a
general algorithm, the Grover algorithm and the Shor's algorithm are
estimated with the optimistic assumptions, in which the only restriction is
from the quantum control. Thus, in Fig.\ref{algorithms} it could be found
that the practise of quantum computation heavily depends on the
sophisticated quantum algorithms and arbitrary quantum operations on about
several tens qubits is already inaccessible even in principle. This handicap
on the quantum computation stands when the quantum computation is carried
out by tandem elementary gates under the quantum control.

\section{Conclusion}

In this paper we present a universal description for the quantum control
based on the quantized controller. We discovered the complementarity about
the competition between the controllability and the control induced quantum
decoherence in the view of quantum measurement. Starting with an
exactly-soluble example, a general model of quantum control is proposed to
describe this novel complementarity or a new type of uncertainty relation.
Our investigations show that it is possible to realize the decoherence free
quantum controls only with some special phases at the finite energy scale
and in finite time. If the parameters of phase is to be determined by the
initial state of the controller, then there exists a low bound for the
systematic errors resulted from the decoherence cause by the quantum control
itself.

The above arguments also show that the decoherences from the quantum control
are different from those induced by the environments through the unwanted
interactions. This is because the negative influence of the controller
happens in the quantum control process itself. If one eliminates this
influence out and out, the positive role of the quantum control would perish
together. Therefore, for quantum computing, these kinds of errors induced by
the control itself can not be overcame totally by the conventional error
management protocols \cite{ECCs}. At least, it has not been proved that the
control induced decoherences can also be conquered efficiently by the
well-estabilished error management protocols. The better method to solve
this problem is to optimize the control operations when the target of
control is given. Without doubt, this is an open question to challenge for
the physical implementation of quantum computing as well as the other
protocols of quantum information processing.

\label{sec:D&C}

This work was supported by CNSF (Grant Nos. 90203018, 90303023, 10474104 and
60433050. It is also funded by the National Fundamental Research Program of
China with Nos. 2001CB309310 and 2005CB724508.


\begin{thebibliography}{99}
\bibitem{q-control} A. Doherty, J. Doyle, H. Mabuchi, K. Jacobs, and S.
Habib, \textit{Robust Control in the Quantum Domain,} Proceedings, 39th IEEE
Conference on Decision and Control (Sydney, December 2000), quant-ph/0105018.

\bibitem{q-inform} M. Nielsen and I. Chuang, \textit{Quantum Computation and
Quantum Information}, (Cambridge University Press, 2000).

\bibitem{SQL} V. Braginsky and F.A. Khalili, \textit{Quantum Measurement},
(Cambridge, 1992); V. Braginsky and F. A. Khalili, Rev. Mod. Phys, \textbf{68%
}, p1 (1996).

\bibitem{Lloyd2000} S. Lloyd, Nature (London) \textbf{406}, p1047 (2000).

\bibitem{Ozawa2002} M. Ozawa, Phys. Rev. Lett. \textbf{89}, 057902 (2002).

\bibitem{Julio2002b} J. Gea-Banacloche, Phys. Rev. Lett. \textbf{89}, 217901
(2002); J. Gea-Banacloche, eprint, quant-ph/0209065 (2002).

\bibitem{q-opt} D.F. Walls and G.J. Milburn,\textit{Quantum Optics},
(Springer-Verlag, 1994).

\bibitem{sun95} C.P. Sun, X.X. Yi and X.J. Liu, Fortschr. Phys. \textbf{43}
(7): p585-612 (1995).

\bibitem{sun91} C.P. Sun and Q. Xiao , Commun. Theor. Phys. \textbf{16},
p359-362 (1991).

\bibitem{heisenberg} W. Heisenberg, \textit{Physical Principles of the
Quantum Theory}, (Dover Publications, 1930).

\bibitem{Aha} A. Stern,Y.Aharonov,Y.Imry, Phys. Rev. A \textbf{{41}%
,3436(1990) }

\bibitem{zls} P. Zhang, X.F. Liu and C.P.Sun, Phys. Rev. A \textbf{66},
042104 (2002).

\bibitem{sun93} C.P. Sun Phys. Rev. A \textbf{48}, p898 (1993); Chinese J.
Phys. \textbf{32}, p7 (1994).

\bibitem{Barenco} A. Barenco \textit{et al.}, Phys. Rev. A \textbf{52},
p3457 (1995).

\bibitem{ECCs} Andrew M. Steane, Phys. Rev. A \textbf{68}, 042322 (2003).
\end{thebibliography}
\end{document}